\documentclass[sn-mathphys-num]{sn-jnl}% Math and Physical Sciences Numbered Reference Style 
\usepackage{graphicx}%
\usepackage{multirow}%
\usepackage{amsmath,amssymb,amsfonts}%
\usepackage{amsthm}%
\usepackage{mathrsfs}%
\usepackage[title]{appendix}%
\usepackage{xcolor}%
\usepackage{textcomp}%
\usepackage{manyfoot}%
\usepackage{booktabs}%
\usepackage{algorithm}%
\usepackage{algorithmicx}%
\usepackage{algpseudocode}%
\usepackage{listings}%
\theoremstyle{thmstyleone}%
\theoremstyle{thmstyletwo}%

\theoremstyle{thmstylethree}%

\begin{document}

\title[Article Title]{Anomalous transport models for fluid classification: insights from an experimentally driven approach}


\author[1,2,6]{\fnm{Sara} \sur{Bernardi}}\email{sara.bernardi@cnr.it}

\author[3]{\fnm{Paolo} \sur{Begnamino}}\email{p.begnamino@eltekgroup.it}

\author[3]{\fnm{Marco} \sur{Pizzi}}\email{m.pizzi@eltekgroup.it}

\author*[1,4,5]{\fnm{Lamberto} \sur{Rondoni}}\email{lamberto.rondoni@polito.it}

\affil[1]{\orgdiv{Department of Mathematical Sciences}, \orgname{Politecnico di Torino}, \orgaddress{\street{Corso Duca degli Abruzzi 24}, \postcode{10129} \state{Torino}, \country{Italy}}}

\affil[2]{\orgdiv{Institute of Atmospheric Sciences and Climate}, \orgname{National  Research Council of Italy}, \orgaddress{\street{Corso Fiume 4}, \postcode{10133} \state{Torino}, \country{Italy}}}

\affil[3]{\orgdiv{Research Department}, \orgname{ELTEK S.p.A.}, \orgaddress{\street{Strada Valenza 5/A},  \postcode{15033} \state{Casale Monferrato, AL}, \country{Italy}}}

\affil[4]{\orgdiv{INFN}, \orgname{Sezione di Torino}, \orgaddress{\street{Via P. Giuria 1}, \postcode{10125} \state{Torino}, \country{Italy}}}

\affil[5]{ORCID:  0000-0002-4223-6279}

\affil[6]{ORCID:  0000-0002-3232-1664}


\abstract{In recent years, research and development in nanoscale science and technology have grown significantly, with electrical transport playing a key role. A natural challenge for its description is to shed light on anomalous behaviours observed in a variety of low-dimensional systems. We use a synergistic combination of experimental and mathematical modelling to explore the  transport properties of the electrical discharge observed within a micro-gap based sensor immersed in fluids with different insulating properties. 
Data from laboratory experiments are collected and used to inform and calibrate four mathematical models that comprise partial differential equations describing different kinds of transport, including anomalous diffusion: the Gaussian Model with Time Dependent Diffusion Coefficient, the Porous Medium Equation, the Kardar-Parisi-Zhang Equation and the Telegrapher Equation. 
Performance analysis of the models through data fitting reveals that the Gaussian Model with a Time-Dependent Diffusion Coefficient most effectively describes the observed phenomena. This model proves particularly valuable in characterizing the transport properties of electrical discharges when the micro-electrodes are immersed in a wide range of  insulating as well as conductive fluids. Indeed, it can suitably reproduce a range of behaviours spanning from clogging to bursts, allowing accurate and quite general fluid classification.
Finally, we apply the data-driven mathematical modeling approach to ethanol-water mixtures. The results show the model's potential for accurate prediction, making it a promising method for analyzing and classifying fluids with unknown insulating properties.
}

\keywords{Nanotechnology, Anomalous diffusion, PDE calibration, Voltage Discharge}



\maketitle

\section{Introduction}\label{sec1}

%classificare le proprietà dei fluidi in nanotecnologia è importante.
%a tale scopo si lavora sulle proprietà elettriche.
%i sensori sviluppati spesso lavorano ad alti campi elettrici.

The measurement of electrical parameters is a common technique used for the characterization of matter properties. Several techniques have been developed for this purpose, and most of them are based on the measurement of current and voltage in different conditions \cite{raju2017dielectrics}. 
The sensor described in the present study is intended to characterize fluids under relatively high fields, close to the breakdown field of the material. 
%The sensor described in the present study is intended to characterize matter under relatively high fields, close to the breakdown voltage of the material. 
For insulators like polymers, the typical dielectric strength values are of the order of some kV/mm. 
For insulating liquids used in high voltage applications the values can be in the range $5-40$ kV/mm or higher. During operation the insulating fluid can be contaminated or degraded, loosing its properties. For safety and efficiency reasons it is crucial to maintain the insulating fluid properties under 
control. Normally used bench techniques are not suitable for miniaturized on-board sensors, requiring high voltages to be operated, with obvious safety and cost related issues. The idea underlying the sensor used in this study is that reducing the distance between electrodes it is possible to obtain very high fields, even at low voltage. Scaling of the electrodes distance allows the measurement of current flow at low voltage. In order to avoid long lasting discharges that could damage the sensor itself, a finite charge is applied, using a capacitor as charge source, as described in the next session. 
The flow of the electric charge in the media between electrodes can follow different regimes, not only reducible to ohmic behavior but including different types of anomalous diffusion.  
In the first scenario, the mean
square displacement is characterized by a time-linear dependence, i.e. $\langle |x(t)-x(0)|^2 \rangle \sim t$,
which is typical
of a Markovian process, and the dynamics of the system is described by the diffusion equation. 
In contrast, the process exhibits \emph{anomalous transport} when it is characterized by a different time dependence for
the mean square displacement, i.e. $\langle |x(t)-x(0)|^2 \rangle \sim t^\gamma$, providing evidence of the non-Markovian nature of the process, \cite{metzler2000random, evangelista2018fractional}. In fact, anomalous electric response is found in several systems, such as fractal electrodes \cite{kumar2009theory},
nanostructured iridium oxide \cite{sunde2010impedance}, and water \cite{batalioto2010dielectric,lenzi2011anomalous}. Several approaches have been
proposed to mathematically describe
these anomalous responses. %
Microscopic models focus on the evolution of systems made of a certan number of particles, and include molecular dynamics techniques \cite{lepri2003thermal,dhar2008heat}, continuous-time random walk, Lévy flights and fractional brownian motion.
While microscopic models gain important insight into the collective behavior starting from the dynamics of the constituting particles and their statistical properties, but at quite high and often prohibitive computational costs, 
macroscopic models offer a simpler and computationally much cheaper framework for understanding how these behaviors scale up to affect transport processes at the macroscopic level.
Specifically, macroscopic models describe the time evolution of particle densities, including different nonlinear diffusion models and fractional diffusion equations.
Fundamental reviews can be found in \cite{metzler2000random}.

In this study, we focus on macroscopic models, specifically four partial differential equations (PDEs): (i)
a generalization
of the Gaussian model with time-dependent diffusion
coefficient, (ii) the Porous Medium Equation, (iii) the Kardar-Parisi-Zhang Equation, and (iv) the Telegrapher Equation.
In order to inform the mathematical models, specific experiments are carried out to explore the transport properties of the elecrical discharge observed within the micro-gap based sensor immersed in fluids with different insulator properties. %Data from laboratory experiments are used to inform and calibrate four mathematical models exhitibing anomalous diffusion.
Even if the sensor was initially designed to measure the properties of insulating fluids, it has been then used also with different kinds of conducting fluids. 
Through data fitting, each fluid is then characterized by the optimal parameters which reveal the transport properties of the process and represent a signature of the material where the discharge at the microscale is measured.
Indeed, the positioning of different materials in a parameter space is a powerful tool for the development of a new class of sensors, where a relatively simple hardware can supply deep understanding of the transport phenomena.  
The rest of the paper is organized as follows. 
In Section \ref{sec2}, the experimental setup is described with the characteristic of the micro-gap sensor used to collect experimental data. Section \ref{sec_models} introduces the continuous models for anomalous transport mentioned above. In Section \ref{results}, we perform the models calibration based on experimental data which allows us to rank the models in terms of fitting performance and identify the TdDC model as the most effective model in describing the observed phenomenon. We then apply the fitting procedure to fluid classification, using both fluids with well known insulating and conductive properties as well as ethanol-water mixture. We conclude with a discussion and suggestions for future investigation.

\section{Motivating experiment: nanotechnology-based sensors}\label{sec2}
\begin{figure}[h!]
    \centering
\includegraphics[width=0.95\textwidth]{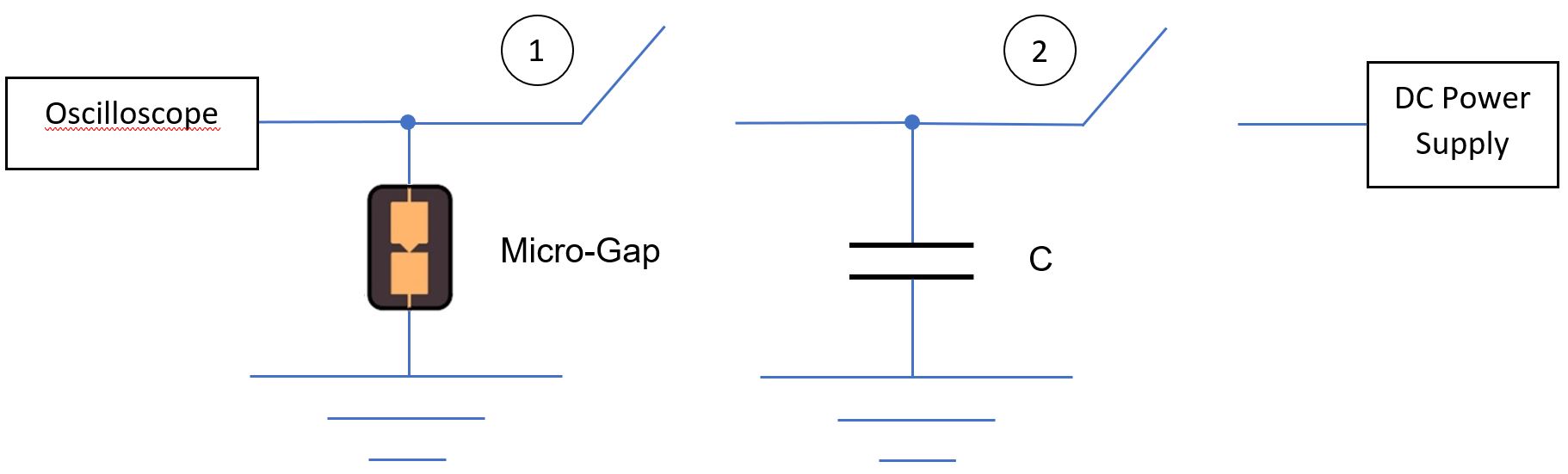}
\caption{Functioning Blok Diagram. A capacitor is fully charged from power supply (switch 1 open, 2 closed), then it is connected to the Micro-Gap that allow electrical discharge (switch 1 closed and 2 open)). The sensor allows to directly detect a possible dacay of the dielectric strength of an insulating fluid.}
    \label{fig:sensore}
\end{figure}

The core of the experimental set up is the microgap: it is formed by a couple of microelectrodes fabricated by photolithography at a distance of few micrometers. Different distances, materials and versions have been tested but a typical microgap is depicted in Figure \ref{fig:sensore} (see also Figure \ref{fig:microgap}A).  
The minimum distance between the electrodes is of $1.5$ microns. The microgap is wired to a switch that can connect the microgap to a $100$ nF capacitor (see n. $1$ in Figure \ref{fig:sensore}). The capacitor is connected by a second switch (n. $2$ in Figure \ref{fig:sensore}) to a DC power supply (model EA-PS 5040-40 A). When the capacitor is under power supply charge, it is disconnected from the microgap (switch n. 2 closed and n. 1 open). When the contact between the power supply and the capacitor is open, the capacitor can be connected to the microgap. The voltage on the microgap is measured by an oscilloscope (model RS PRO IDS-1104B).

    If the microgap is immersed in a virtually perfect insulator, the discharge time of the capacitor is virtually infinite. In real media a current can flow between the microelectrodes even at finite times. The current flow in the microgap can then be used to characterize the fluid, as it will be clear in the next sessions. 

\section{Anomalous diffusion models for the discharge between micro-electrodes}\label{sec_models}

\begin{figure}[h!]
    \centering
\includegraphics[width=0.7\textwidth]{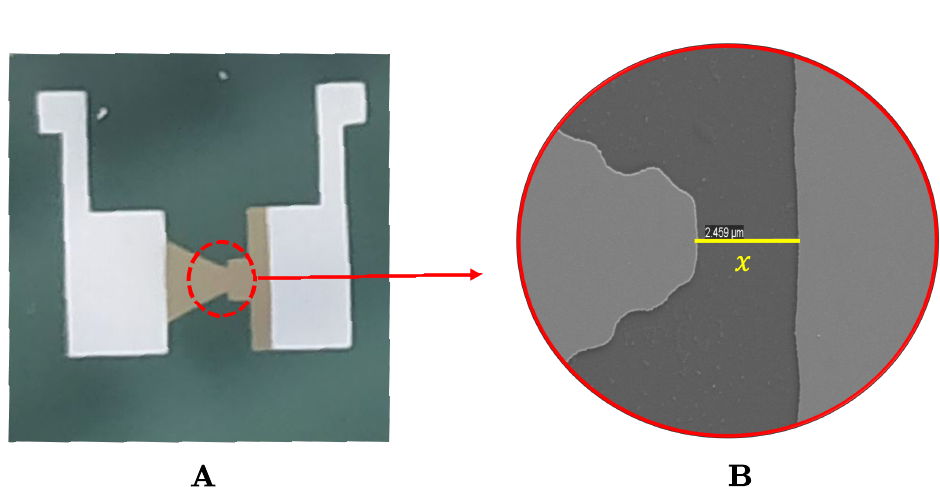}
    \caption{(A) Microgap made of a metal thin film, typically Cr 100 nm thick, deposited on an insulating layer (SiO$_2$ or Al$_2$O$_3$)  with electric contacts (Al or Au, 150 nm thick). (B) Detail of the part of the device where the distance between electrodes is minimum. The spatial domain $\Omega$ is set to represent the micro-gap extension.}
    \label{fig:microgap}
\end{figure}

The above-described experimental setting is modeled in terms of the voltage $u(x,t)$ at position $x$ in the one-dimensional domain $\Omega=[0,1]$ representing the micro-gap extension (see Figure \ref{fig:microgap}), for all times $t$ in a given interval $[0, T]$. 
We fix proper initial and boundary conditions, that will be then coupled with different evolution laws describing anomalous transport within the micro-gap.
Specifically, we set an unbalanced initial voltage configuration, represented by a reflecting boundary condition at $x=0$ and absorbing boundary condition at $x=1$. In mathematical terms, we write: 

\begin{equation}
u(x,0) = U_0 (1-x)^n , 
%\label{IC} 
\quad
\frac{\partial u(0,t)}{\partial x} = 0 \, , %\label{BC_l} 
\quad
u(1,t) = 0 
\label{BC_r}
\end{equation}
where $U_0 > 0$ is the initial charge on the left electrode.
We then consider different evolution laws such that the mean square displacement of charges, $\mathbb{E}(X^2(t))$ is asymptotically proportional to $t^\gamma$, with $0 \leq \gamma \leq 2$.

\paragraph{Gaussian model with Time Dependent Diffusion Coefficient}

\begin{figure}[H]
    \centering
\includegraphics[width=1\textwidth]{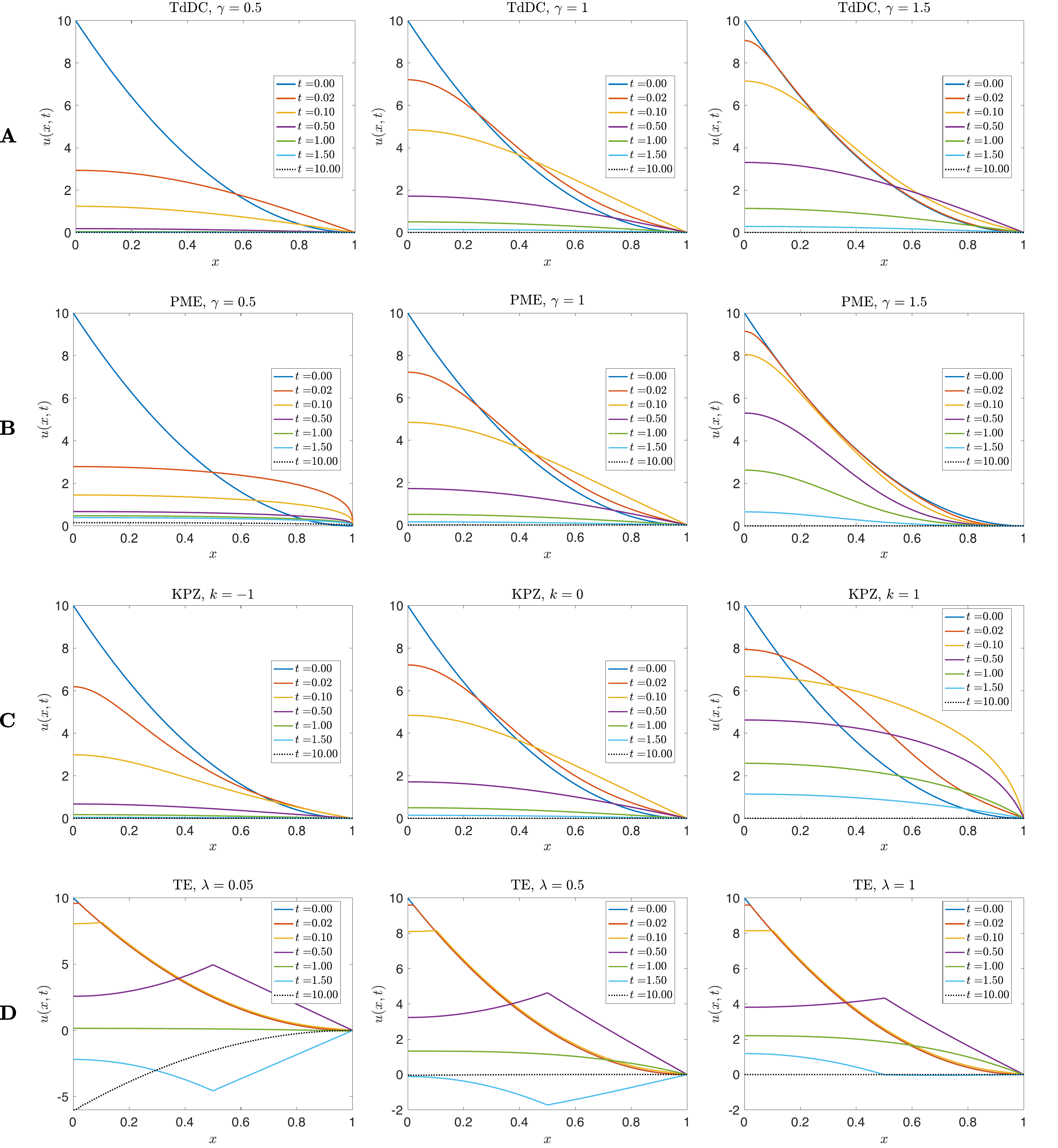}
    \caption{Time evolution of models solution. (A) TdDC model, for different values of $\gamma$ and $D=1$. (B) PME model, for different values of $\gamma$ and $D=1$. (C) KPZ model, for different values of $k$ and $D=1$. (D) TE model, for different values of $\lambda$ and $s=1$. Parameters in the initial condition are set as $U_0=10$, $n=2$.}
    \label{fig1}
\end{figure}

The Gaussian model with Time Dependent Diffusion Coefficient (TdDC) derives from a generalization of the Diffusion equation which includes a general spreading of the density, i.e. anomalous diffusive transport (for a detailed derivation, see \cite{bernardi2024anomalous}). 
It reads as
\begin{equation}
\frac{\partial u}{\partial t}=D t^{\gamma-1} \frac{\partial^2 u}{\partial x^2}.
\label{G_PDE}
\end{equation}
Specifically, it introduces a temporal dependence of the diffusion coefficient $\tilde{D}=D t^{\gamma-1}$ which describes normal diffusion for $\gamma=1$ and anomalous diffusion for $\gamma \neq 1$, see the numerical solution in Figure \ref{fig1}A. In the second case, $\gamma < 1$ implies clogging, {\em i.e.}\ slowing down in time of the transport, that may even stop altogether; while $\gamma > 1$ means that this rate rapidly increases, allowing a burst or a sort of micro-lightning.

\paragraph{Porous Medium Equation}

The Porous Medium Equation (PME) describes density-dependent diffusivity in porous media
\begin{equation}
\frac{\partial u}{\partial t}= D\frac{\partial^2 u^m}{\partial x^2} =
\frac{\partial }{\partial x} \left( D
m u^{m-1} \frac{\partial u}{\partial x} 
\right), \quad m>1,
\label{PME}
\end{equation}
which has been proved to be related with anomalous transport both from theoretical \cite{voigtmann2009double, dos2019analytic} and experimental \cite{fatin2004size} studies.
Moreover, simple algebra provides an explicit relation between the exponent appearing in the PME and the anomalous transport feature of its solution, i.e. $m=(2-\gamma)/{\gamma}$ (for explicit derivation, see \cite{bernardi2024anomalous}).
Equation \eqref{PME} thus describes normal diffusion for $\gamma = 1$ (i.e. $m = 1$) and anomalous diffusion for $\gamma \neq 1$ (i.e. $m \neq 1$), see the solution profiles in Figure \ref{fig1}B.

\paragraph{Kardar-Parisi-Zhang Equation}

The Kardar-Parisi-Zhang equation was first introduced to recapitulate universal aspects of growing interfaces \cite{kardar1986dynamic}, and later connected to anomalous transport phenomena \cite{lepri2016heat}. It reads as
\begin{align}
&\frac{\partial u}{\partial t} = \frac{k}{2} \left(\frac{\partial u}{\partial x}\right)^2+D \,\frac{\partial^2 u}{\partial x^2} + \eta,
\label{KPZ_eq}
\end{align}
where $k$ and $D$ are positive constants, and $\eta$ is white Gaussian noise with average $ \langle \eta(x,t) \rangle =0$ and 
instantaneous decay of correlations: $\langle \eta(x,t) \eta(x',t') \rangle=2D \, \delta(x-x') \, \delta(t-t')$. 
It has been used to describe different physical phenomena such as turbulent liquid crystals \cite{takeuchi2010universal}, crystal growth on a thin film \cite{wang2003universality}, bacteria colony growth \cite{wakita1997self,matsushita1998interface} and burning fronts \cite{miettinen2005experimental}.
We here neglect the noise term, i.e. we set $\eta(x,t) = 0$, because our tests revealed that noise is minimal or totally absent in our experiments. Figure \ref{fig1}C show the effect of varying the parameter $k$ on the resulting solution.

\paragraph{Telegrapher Model}
The Telegrapher model describes a
finite-velocity diffusion process. It has been derived in the context of electrodynamic theory \cite{heaviside2011electrical}, and later proposed as a model for the transport in turbulent diffusion \cite{taylor1922diffusion} as well as for the lightning processes \cite{rakov2003lightning}. It is formulated as

\begin{align*}
&\frac{\partial^2 u}{\partial t^2}+2\lambda \frac{\partial u}{\partial t}-s^2 \frac{\partial^2 u}{\partial x^2}=0,
\end{align*}
where $s^2$ and $\lambda$ denote the diffusion and damping coefficients. For more details on derivation and application of the telegrapher’s equation, we refer to the literature \cite{weiss2002some, joseph1990addendum}.
%The Telegrapher equation can be also considered as a particular case of a spatio-temporally coupled Lévy walk model with exponential waiting time probability density \cite{zumofen1993scale, klafter1994levy,zaburdaev2015levy}. Furthermore, it has  
Figures \ref{fig1}D show how variations in the damping coefficient $\lambda$ affect the resulting solution.

\section{Results}\label{results}
In this section, we inform and calibrate the PDE models introduced above to shed light on experimental time series data on voltage discharge observed between the sensor micro-electrodes.

\subsection{Models calibration based on experimental data}

\begin{figure}[H]
    \centering
\includegraphics[width=0.8\textwidth]{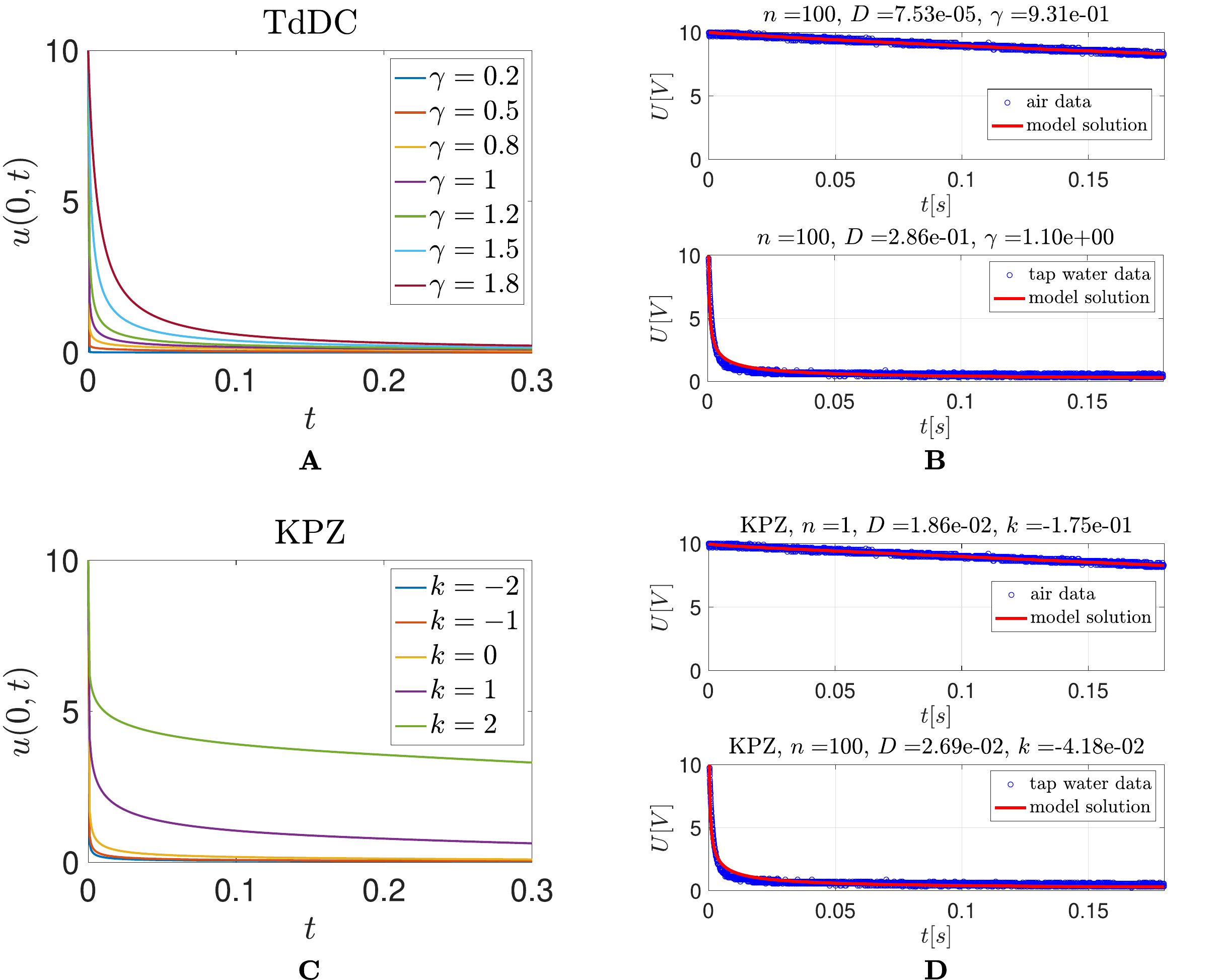}
    \caption{(A-C) TdDC and KPZ solutions at $x=0$. Models parameters are set as $D=1$, $n=100$, $U_0=10$. (B-D) fitted solutions obtained through model calibration to experimental data.}
    \label{fig3}
\end{figure}

\begin{figure}[H]
    \centering
\includegraphics[width=0.8\textwidth]{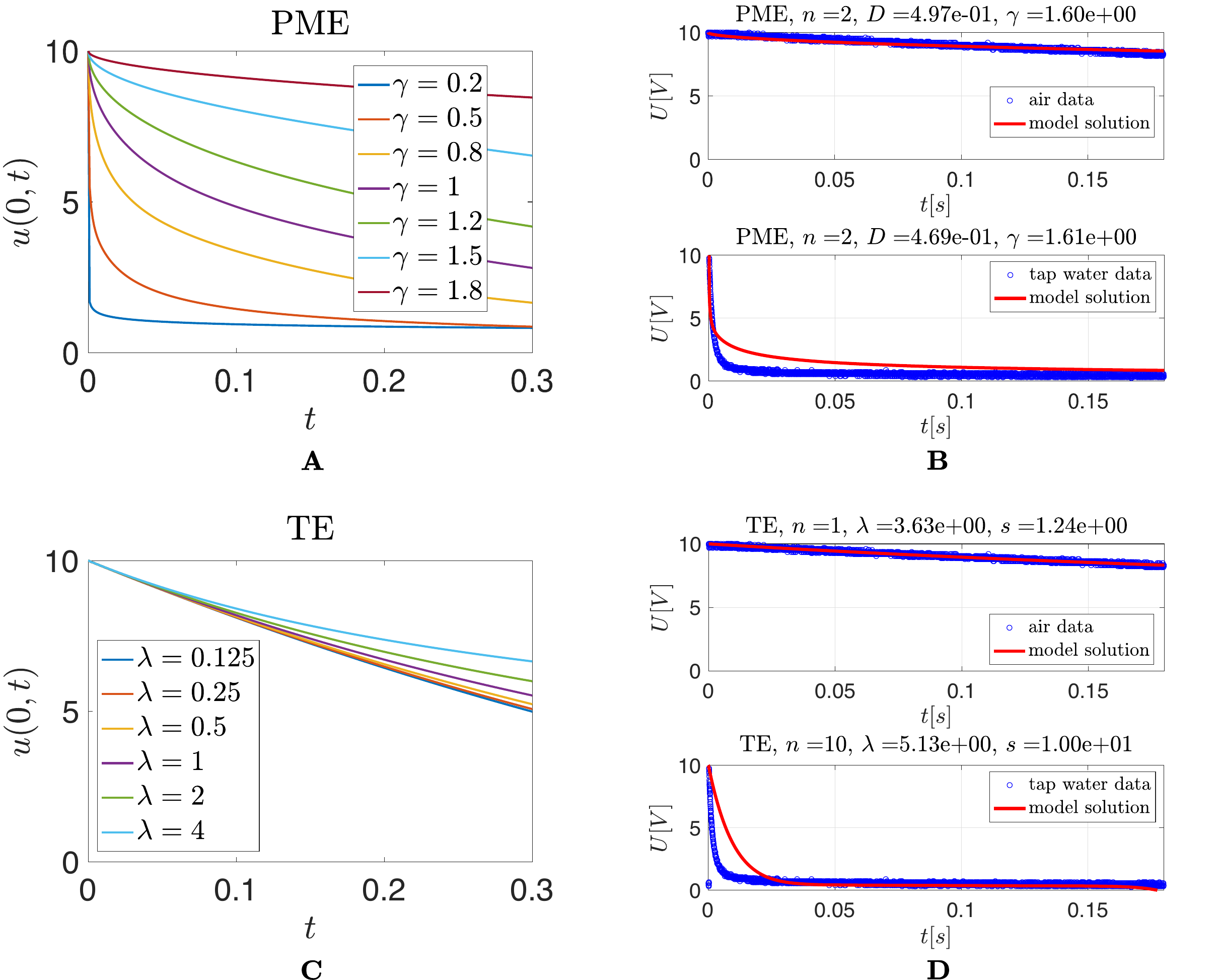}
    \caption{(A-C) PME and TE solutions at $x=0$. Models parameters are set as: $D=1$, $n=100$, $U_0=10$ (PME); $s=1$, $n=2$, $U_0=10$ (TE). (B-D) Fitted solutions obtained through model calibration to experimental data.}
    \label{fig4}
\end{figure}

Experimental time series data on voltage discharge between the sensor micro-electrodes are collected and used to investigate which model better describes the physical phenomenon. We assume that the voltage measurement is represented by the model's solution at the left boundary of the domain, $u(0,t)$. The other electrode is grounded, $u(1,t)=0$.
Model calibration is then performed using a classical Bayesian method, leveraging the built-in MATLAB routine \texttt{lsqcurvefit}, which is based on the least-squares method.
Specifically, given the experimental observations $(t_i,V_i)$, for $i=1,...,M$, the optimal parameter set is obtained by minimising the sum of squared residuals 
\begin{equation}\label{R}
R=\min_p ||u_p(0,t)-V||_2^2=\min_p \sum_{i=1}^M (u_p(0,t_i)-V_i)^2,
\end{equation}
where $u_p$ is the model solution obtained for the parameter set $p$.
Figures \ref{fig3} and \ref{fig4} compare the time evolution of $u(0,t)$ for different model parameter values (panels A-C) and the fitted solutions obtained by adapting the models to the voltage data observed when the micro-electrodes are immersed in two illustrative fluids with opposite conductive properties, i.e., air and tap water (panels B-D). %Indeed, we aim to find a model capable of describing both slow and fast electric discharge. 
Figures \ref{fig3}B, \ref{fig3}D, \ref{fig4}B and \ref{fig4}D show that all the PDE models can qualitatively describe electrical discharge.

\begin{figure}[H]
    \centering
\includegraphics[width=1\textwidth]{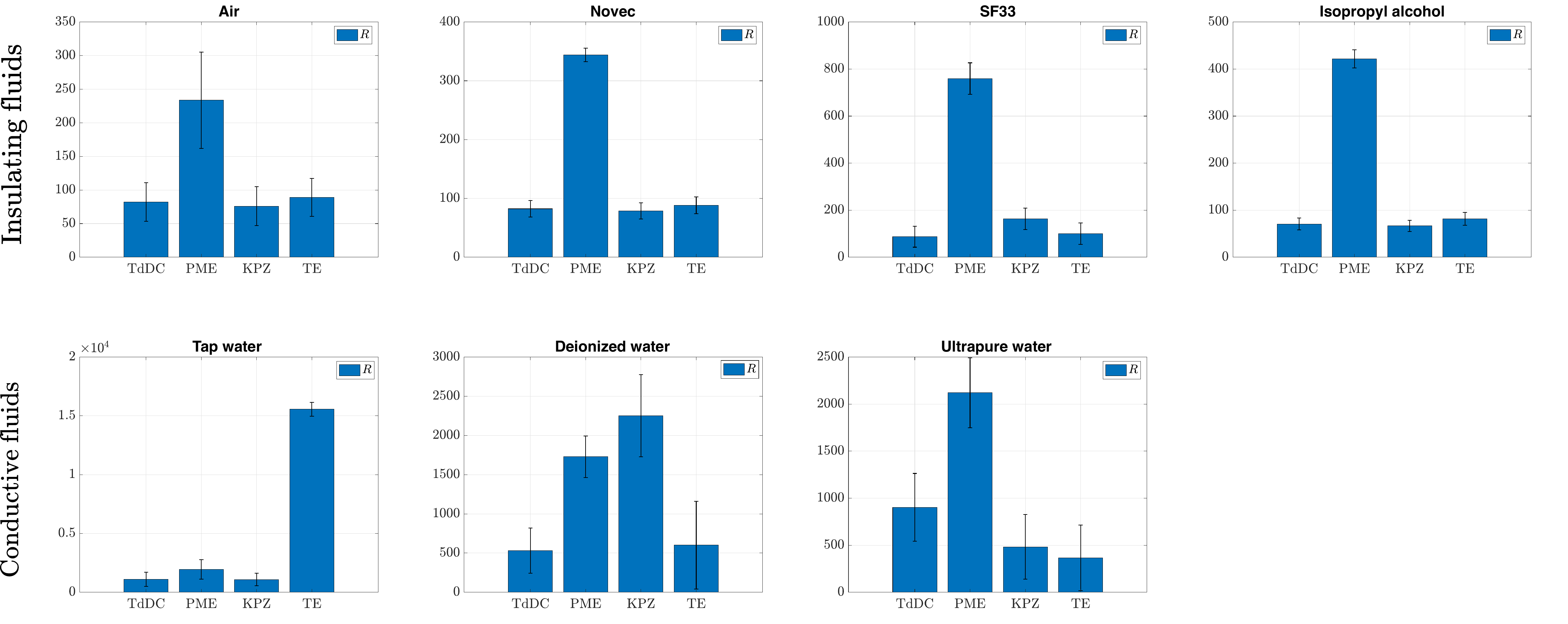}
    \caption{Performance comparison of the fitting procedure applied to the PDE models, with the steepness parameter $n$ of the initial condition chosen among three values ($n=1$, $n=10$ and $n=100$). For each fluid, the mean and standard deviation of the squared 2-norm of the residuals, obtained from the optimal parameter sets, are evaluated over a set of 10 experimental replicates. }
    \label{ranking}
\end{figure}

\begin{figure}[H]
    \centering
\includegraphics[width=1\textwidth]{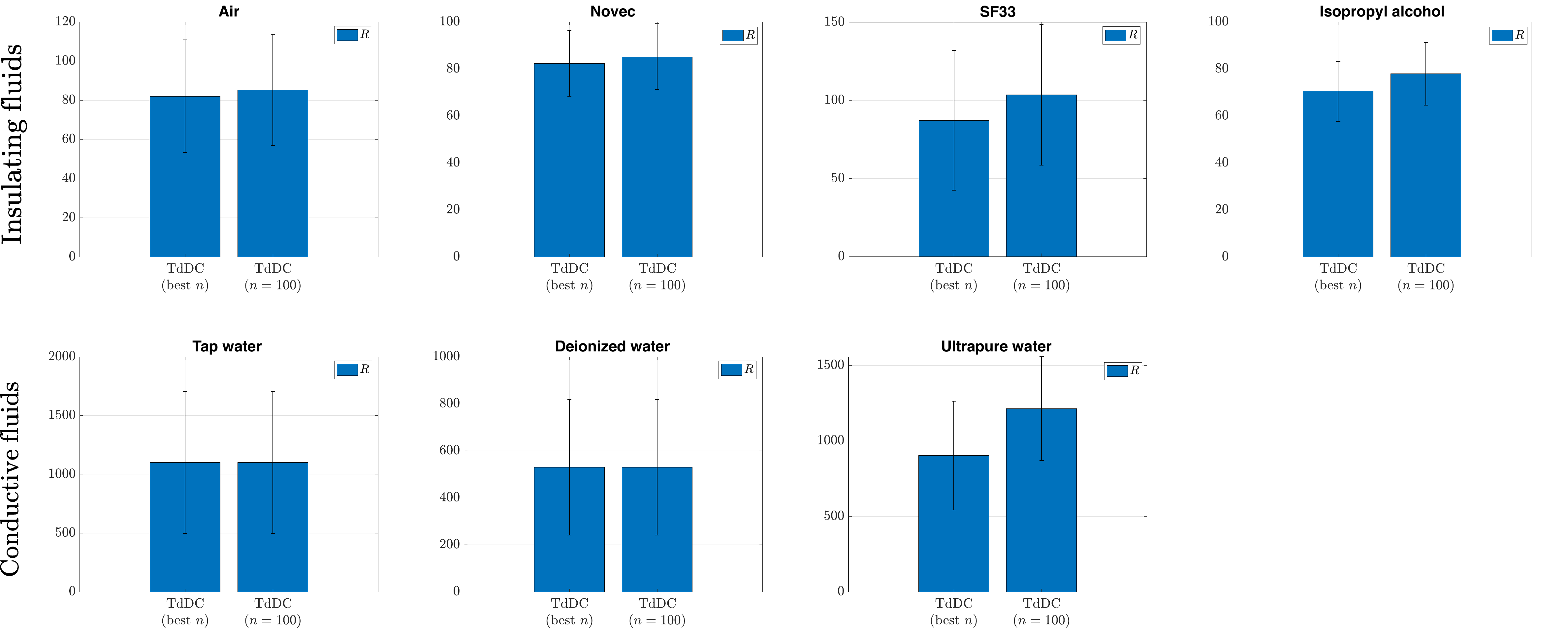}
    \caption{Performance comparison of the fitting procedure applied to the TdDC model, with the steepness parameter $n$ of the initial condition chosen among three values ($n=1$, $n=10$ and $n=100$), and with $n$ fixed ($n=100$). For each fluid, the mean and standard deviation of the squared 2-norm of the residuals, obtained from the optimal parameter sets, are evaluated over a set of 10 experimental replicates. }
    \label{ranking_TdDC}
\end{figure}

To determine which model best replicates the phenomenon, we evaluate the squared 2-norm of the residual obtained for the optimal parameter set, i.e. $R$ defined in Eq. \eqref{R}, to quantify the goodness of the fit for the four PDE models. For this purpose, we use measurements of electrical discharge observed when a 12 V electrical voltage is applied to micro-electrodes immersed in fluids with varying conductive properties. Specifically, both insulating fluids, i.e. air, Novec, SF33, and isopropyl alcohol, as well as conductive fluids, including tap water, deionized water, and ultrapure water, are used. The mean and standard deviation of 
$R$, evaluated over a set of 10 experimental replicates for each fluid, are shown in Figure \ref{ranking}. Specifically, for each of the 10 replicates, $R$ is evaluated as the minimum value obtained by the fitting procedure upon the choice between three different values of the steepness $n$ of the initial condition, i.e. $n=1, n=10, n=100$. 
However, the option $n=100$ is excluded for the TE due to numerical instability issues.
Overall, the best fit performance is obtained for the Gaussian model with Time dependent Diffusion Coefficient (TdDC). Moreover, Figure \ref{ranking_TdDC} shows that fixing 
$n=100$ does not significantly affect the performance of the fitting procedure. Because we also need a steep initial profile, to represent situation before the discharge,
we will adopt this choice, and the TdDC model, in the following sections.

\subsection{Fluid classification}

The calibration procedure illustrated above allows us to identify each time series by a set of optimal parameters $(\gamma, D)$ for the TdDC model, which reveal the transport properties of the observed electrical discharge. Moreover, the optimal parameter space resulting from the data fitting highlights regions that can be used to classify fluids with different insulation properties.

%specificare Equazione e che n = 100.
\paragraph{Conductive and insulating fluids}

\begin{figure}[H]
    \centering
\includegraphics[width=1\textwidth]{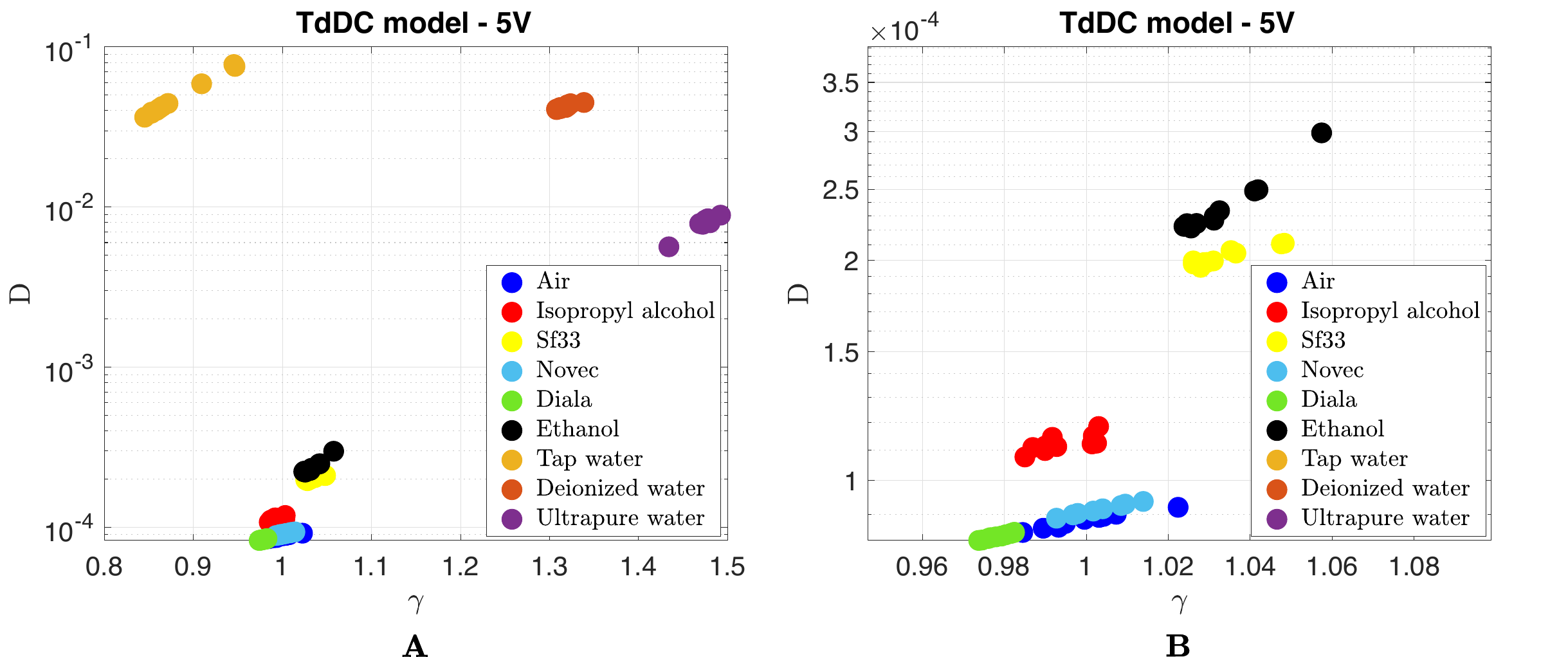}
    \caption{Optimal parameters obtained for conductive and insulating fluids, obtained from fitting the data on the voltage discharge observed when a 5 V electrical voltage to the micro-electrodes is applied.}
    \label{Ris_5V}
\end{figure}

We have collected  experimental measurement of electrical discharge observed applying a 5 V electrical voltage to the micro-electrodes immersed in fluids with known insulating properties. Specifically, we used both insulating fluids, including air, Isopropyl alcohol, Sf33, Novec, Diala, Ethanol, and conductive fluids, namely tap water, Deionized water and Ultrapure water. For each fluid, we measured 10 replicates.
The optimal parameter values, represented as coordinates in the parametric space $(\gamma, \log D)$, are shown in Figure \ref{Ris_5V}.

\begin{figure}[H]
    \centering
\includegraphics[width=1\textwidth]{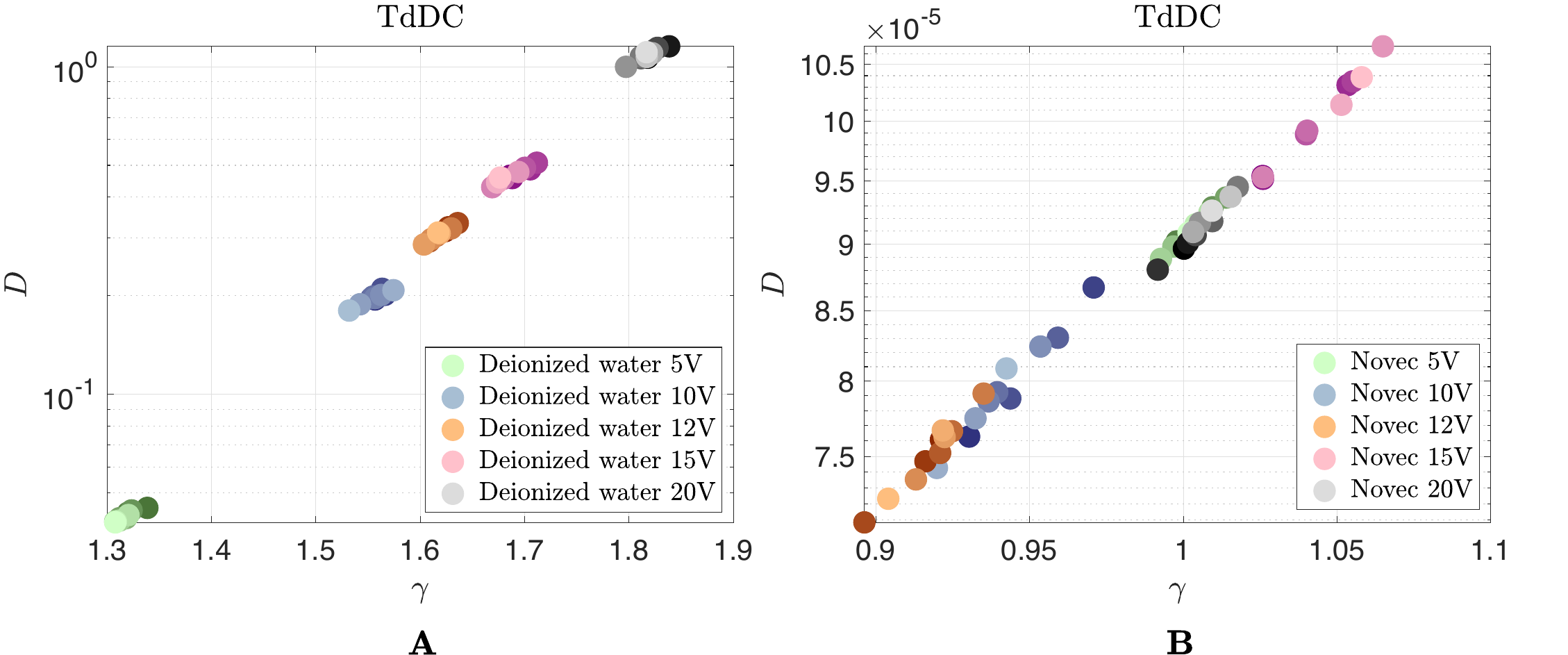}
    \caption{The optimal parameters obtained for (A) Deionized water and (B) Novec from fitting the data on voltage discharge observed when an electrical voltage of 5V, 10V, 12V, 15V and 20V is applied to the micro-electrodes. The successive tests are indicated by a color gradient from darker to lighter shades. This color code does not reveal any temporal ordering of the results. On the other hand, the exponential dependence of $D$ on $\gamma$ for the single fluids and for all of them is evident.}
    \label{Novec_Deion}
\end{figure}

The repeated tests for each fluid are well localized in the parametric space, supporting that the fit results are robust. Notably, the optimal parameters $(\gamma, D)$ exhibit a linear trend within the parametric space.
The same behavior is observed at higher voltage configurations, such as 10V, 12V, 15V, and 20V. Specifically, at higher initial voltages, conductive fluids are associated with higher values of both the optimal $D$ and $\gamma$ parameters (see Figure \ref{Novec_Deion}A). In contrast, insulating fluids do not exhibit significant changes in their position in the parametric space as the initial voltage varies 
(see Figure \ref{Novec_Deion}B). Moreover, the color gradient in Figure \ref{Novec_Deion} shows that the successive trials do not align in an orderly manner along the trend line. 
This disorder indicates that the fluid does not deteriorate as tests are repeated, 
but it is more likely due to variations in the initial conditions, i.e. in the charge of the electrodes at time $t = 0$.

The optimal parameters resulting from the discharge in insulating fluids are all located in a neighbourhood of $(\gamma=1, D=10^{-4})$, see also Table \ref{tab:fluidi_isol}. This reflects a normal diffusion process with a low diffusion coefficient.
Even if the values are quite similar it is worth noting that fluids with very good insulating properties, with dielectric strength higher than 30 KV in standard testing conditions, can be  distinguished from less insulating fluids like SF33. In particular the estimated value of the diffusion parameter of the SF33 is approximately two times the values of Novec or Diala insulating fluids, see Table \ref{tab:fluidi_isol}. The result is very relevant from the point of view of the application because a non destructive technique, operated at low voltage, is predictive regarding the behaviour of the fluid at high voltage. 
The discharge in conductive fluids instead results in greater values of $\gamma$ and $D$, or both, suggesting a super-diffusive process with a higher diffusion coefficient. 
The positioning of the optimal parameters in the parametric space thus suggests a robust method for classifying fluids with different insulation properties. %The correct theoretical description and parameter estimation of the model is enabling the design of a novel class of sensors, as it will be discussed in the next sessions. 

%commento sulla classificazione

\begin{table}[]
    \centering
    \begin{tabular}{c|c|c|c|c}
       Fluid  &  Average Dielectric Strength (kV) & Std. Dev. & $\overline{\gamma}$ & $\overline{D}$\\   \hline
        Novec 7100 & 48.27 & 3.12  & 1.0020 & 9.1033e-05 \\ \hline
        DIALA S4 ZX-I & 38.35 & 7.17 &9.7849e-01 &8.3962e-05\\ \hline
        SF33 & 2.02   & 0.01 &1.0337 &2.0222e-04
        \\ \hline
    \end{tabular}
\caption{
The average dielectric strength and its standard deviation are measured 
and
compared 
with the mean values of the optimal parameters, $\overline{\gamma}$ and $\overline{D}$, evaluated across a set of ten experimental replicates. }
    \label{tab:fluidi_isol}
\end{table}

%

%\begin{figure}[H]
 %   \centering
%\includegraphics[width=1\textwidth]{risultati.pdf}
 %   \caption{Optimal parameters obtained for conductive and insulating fluids, 12 V.}
  %  \label{Ris_12V}
%\end{figure}

\paragraph{Ethanol-water mixtures}

\begin{figure}[H]
    \centering
\includegraphics[width=1\textwidth]{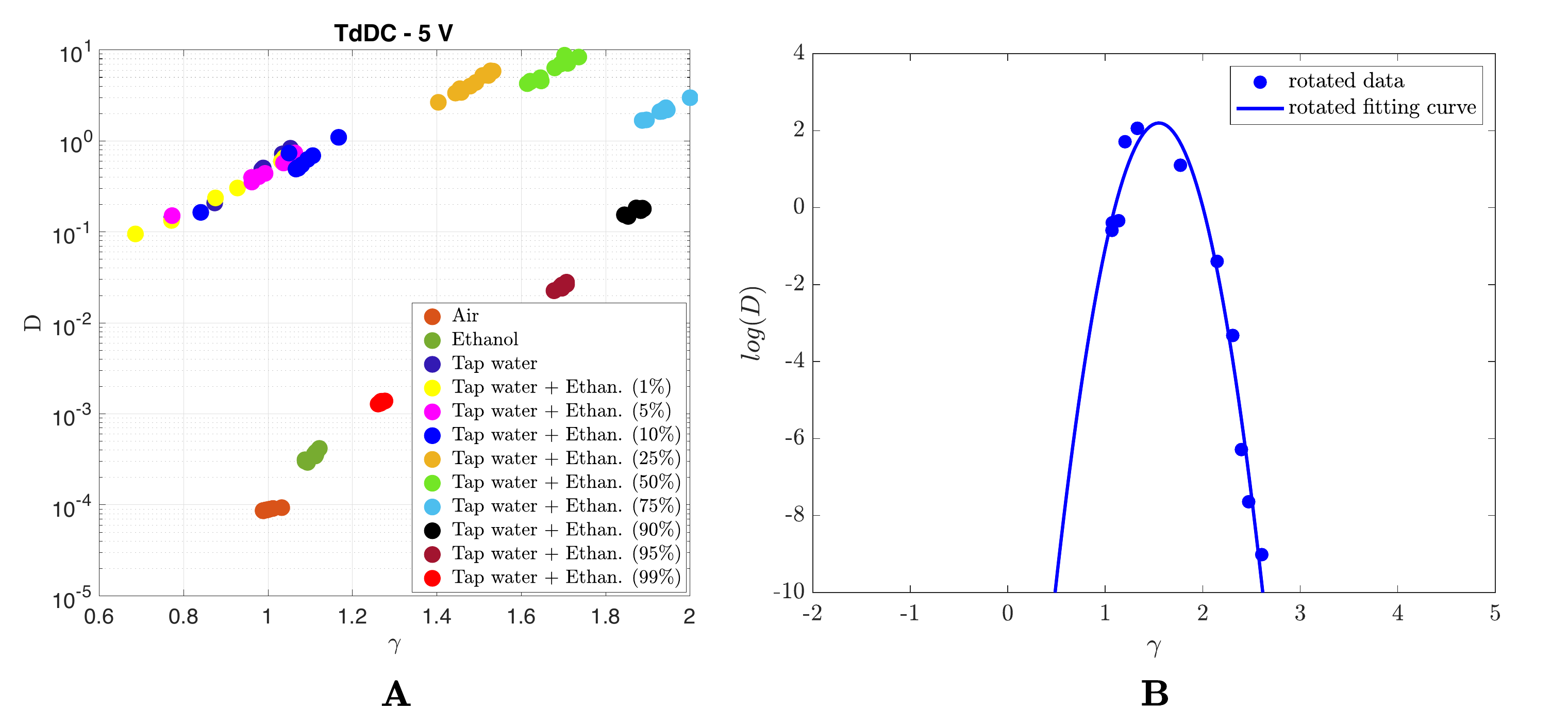}
    \caption{(A) Optimal parameters for mixtures of tap water and ethanol are obtained by fitting data on the voltage discharge observed when a 5 V electrical potential is applied to the micro-electrodes. (B) Centroids of the $(\gamma, \log(D))$ optimal parameters are rotated by 10 degrees (blue points) and fitted to a parabola (blue curve).} %The fitted parabola is then rotated back (red curve) to align with the original data centroids.}
    \label{Ris_miscele_5V}
\end{figure}

The same procedure is applied when the micro-electrodes are immersed in different mixtures of ethanol and tap water, with ethanol concentration varying from $1\%$ to $99\%$. %, specifically different percentages of ethanol ($1\%$, $5\%$, $10\%$, $25\%$, $50\%$, $75\%$, $90\%$, $95\%$, $99\%$) in tap water. 
The resulting optimal parameter sets are represented in the parametric space in Figure \ref{Ris_miscele_5V}A. The mixtures are clearly distinguished in the parametric space, except for those with less than $25\%$ ethanol in water, which are described by similar transport parameters and overlap in the parametric space,
as required for fluids with same conductivity.
Interestingly, the optimal parameters corresponding to mixtures with increasing percentages of ethanol in tap water do not follow the path of minimal distance in the parametric space. Conversely, they follow a specific functional form resembling a tilted parabola.
The same functional trend is observed also for higher voltage initial configuration, i.e. 12 V and 24 V. 
In Figure \ref{Ris_miscele_5V}B, the centroids of the $(\gamma, log(D))$
optimal parameters obtained from the ten repeated tests are fitted with a parabolic curve. Specifically, the fitting curve $p(x)$ is determined by rotating the data by 10 degrees and applying a parabolic fit using MATLAB's built-in \texttt{polyfit} function. The resulting polynomial is given by
$p(x)= -1.0757e+01 x^2 +  3.3358e+01x  -2.3662e+01$.
%, and then rotating the fitted curve back to align with the data centroids.
This straightforward fit holds significant practical relevance, as it can be used where data are missing and, in particular, to predict 
the proportion of an unknown tap water–ethanol mixture, as discussed in the next conclusive section.

%\begin{figure}[H]
%    \centering
%\includegraphics[width=1\textwidth]{risultati_miscelaH2O_eta.eps}
 %   \caption{(A) Optimal parameters obtained for mixtures of tap water and ethanol, obtained from fitting the data on the voltage discharge observed when a 5 V electrical voltage to the micro-electrodes is applied. (B)}
%\label{Ris_fit_miscele_parabola_5V}
%\end{figure}

\section{Discussion}

\begin{figure}[H]
    \centering
\includegraphics[width=0.8\linewidth]{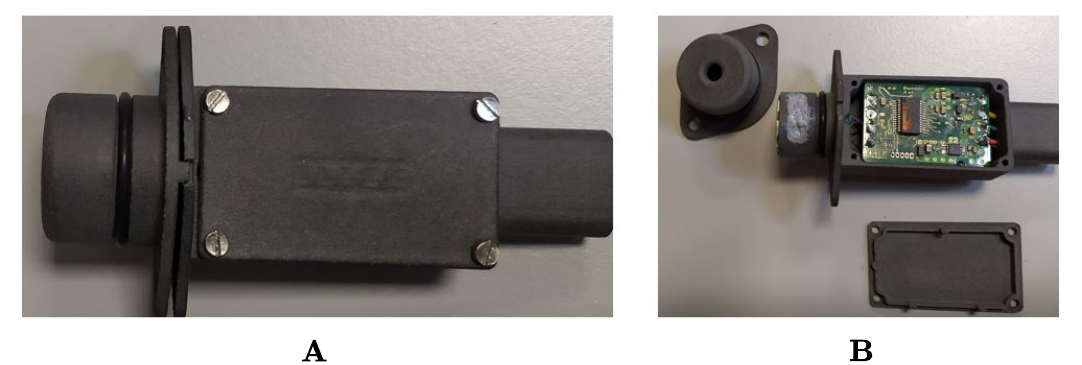}
\caption{(A) Microgap sensor: assembled. (B) Opened, with electronics.}
    \label{fig:microgap_disc}
\end{figure}

Continuous and real-time monitoring of fluid properties is essential for industrial applications, as it enables predictive maintenance of equipments with significant improvement in fluid waste reduction and equipment safety. The class of sensors developed by Eltek used in this study surpasses traditional techniques by employing ``micro-gaps'' as electrodes, allowing it to operate at low voltages (5-12 V), also on small liquid volumes (few microliters are generally sufficient), in a compact miniaturised package that can be adapted to several environments, see Figure \ref{fig:microgap_disc}.
We adopted an experimentally-informed mathematical modeling approach to investigate the anomalous behavior of the electrical discharge observed within the mentioned sensor immersed in fluids with varying insulating properties. From a modeling perspective, we focused on four models consisting of partial differential equations to describe the discharges: the Gaussian Model with 
Time-Dependent Diffusion Coefficient, the Porous Medium Equation, the Kardar-Parisi-Zhang Equation, and the Telegrapher Equation. Data fitting revealed that the Gaussian Model with Time-Dependent Diffusion Coefficient most accurately replicates the experimental dataset. Moreover, the optimal parameters $(D,\gamma)$ not only elucidate the anomalous transport properties of the electrical discharge in the specific fluid but also suggest a novel method for fluid classification that is robust to both replicate measurements and variations in the applied initial voltage. Additionally, we applied the proposed approach to ethanol-water mixtures, demonstrating the potential for accurate predictions on the composition of unknown relative percentages in fluid mixtures. The industrial application of the approach is under evaluation not only for liquid mixtures classification but also for gas and gas-liquid mixtures. For example, the simplest application of the device in gas-liquid mixtures is the rapid measurement of water-air mixtures, acting as a fast humidity sensor.
The described mathematical approach is indeed as simple as powerful and is ready to be applied when a moderate computing power is available. 
Indeed, the simple positioning of the model's optimal parameters $(\gamma, D)$ in the parametric space can uncover differences in the insulative or conductive properties of fluids that conventional techniques may fail to detect. Furthermore, as seen with the specific tap water-ethanol mixture, this approach can be extended for predictive purposes: the fitting results can help in providing insights into the fluid's composition.
The ongoing effort is aimed at the further simplification of the required electronic hardware by the implementation of a low cost miniaturised electronics based on neural networks. 
The training of the networks, and all the related needed power of calculation, will be done in advance on a wide experimental and synthetic data set, generated by the presented model. Once trained, the networks will be implemented on low cost electronics, enabling the production of sensors operating autonomously and not necessarily connected to remote processing units.

\bmhead{Acknowledgements}

SB and LR, as members of GNFM (Gruppo Nazionale per la Fisica Matematica) of INdAM (Istituto Nazionale di
Alta Matematica), acknowledge that this work has been performed under the auspices of GNFM of INdAM. This work is part of the project NODES which has received funding from the MUR - M4C2 1.5 of PNRR funded by the European Union - NextGenerationEU (Grant agreement no. ECS00000036). Thanks to Mauro Zorzetto, Matteo Rondano and Marco Ferragatta of the Mobility Dep. of Eltek, who designed the electronic circuit of the sensor, for the useful discussions on microgap applications. 

\section*{Declarations}

\begin{itemize}
    \item Ethics, Consent to Participate, and Consent to Publish declarations. Not applicable.
    \item Funding. This work is part of the project NODES which has received funding from the MUR - M4C2 1.5 of PNRR funded by the European Union - NextGenerationEU (Grant agreement no. ECS00000036).
    \item Authors’ contributions.  S.B.: conceptualization, formal analysis, investigation, methodology, data curation, software, validation, visualization, writing—original draft, writing—review and editing. P.B.: conceptualization, data curation, investigation, methodology, visualization, writing—review and editing; M.P.: conceptualization, investigation, data curation, methodology, supervision,  writing—review and editing; L.R.: conceptualization, formal analysis, funding acquisition, investigation, methodology, supervision, writing—review and editing.
    \item Competing interest declaration. We declare we have no competing interests. 
\end{itemize}


\begin{appendices}
\section{Numerical methods}
Numerical simulations of the four PDE models are performed on the 1D spatial domain $[0,1]$, with initial and boundary conditions imposed at 
$x=0$ and $x=1$, as specified in Eq. \eqref{BC_r} (main text). The Gaussian Model with Time-Dependent Diffusion Coefficient, the Porous Medium Equation, and the Kardar-Parisi-Zhang Equation are solved using the MATLAB routine \texttt{pdepe}. The Telegrapher Equation is solved by discretizing the spatial derivative, and solving the associated system of two first-order ODEs (for $u$ and its time derivative) using the MATLAB solver routine \texttt{ode45}.




\end{appendices}

\bibliography{sn-article}% common bib file

\end{document}